# Structural peculiarities of ε-Fe$_2$O$_3$ / GaN epitaxial layers unveiled by high-resolution transmission electron microscopy and neutron reflectometry

Sergey M. Suturin[1*], Polina A. Dvortsova[1], Leonid A. Snigirev[1], Victor A. Ukleev[2], Takayasu Hanashima[3], Marcos Rosado[4], Belén Ballesteros[4]

[1]*Division of Solid State Physics, Division of Physics of Dielectrics and Semiconductors, Centre of nanoheterostructure Physics, Ioffe Institute, St. Petersburg, Russia;*

[2]*Laboratory for Neutron Scattering and Imaging (LNS), Paul Scherrer Institute (PSI), Villigen, Switzerland;*

[3]*Neutron Science and Technology Center, CROSS, Tokai 319-1106, Japan;*

[4]*Catalan Institute of Nanoscience and Nanotechnology (ICN2), CSIC and BIST, Campus UAB, Bellaterra, Barcelona, 08193, Spain*

\* Correspondence e-mail: suturin@mail.ioffe.ru

The present paper is dedicated to the structural study of crystallographic peculiarities appearing in epitaxial films of metastable epsilon iron oxide (ε-Fe$_2$O$_3$) grown by pulsed laser deposition onto a semiconductor GaN (0001) substrate. The columnar structure of the nanoscale ε-Fe$_2$O$_3$ films has been for the first time investigated using high resolution electron microscopy (HRTEM) direct space technique complemented by reciprocal space methods of high-energy electron diffraction and color-enhanced HRTEM image Fourier filtering. The study of ε-Fe$_2$O$_3$ / GaN interface formation has been further expanded by carrying out a depth resolved analysis of density and chemical composition by neutron reflectometry and energy-dispersive X-ray spectroscopy. The obtained results shed light onto the properties and the origin of the enigmatic few-nanometer thick low density transition layer residing at the ε-Fe$_2$O$_3$ / GaN interface. A detailed knowledge of the properties of this layer is believed to be highly important for the development of ε-Fe$_2$O$_3$ / GaN heterostructures that can potentially become part of the iron-oxide based ferroic-on-semiconductor devices with room temperature magneto-electric coupling.

# 1. INTRODUCTION

The persistent interest to the nanostructures based on the large family of iron oxides is motivated by the outstanding physical properties of these materials. Having a very simple chemical formula, iron oxides exhibit a vast variety of magnetic phenomena not yet fully explained, such as Vervey transition in magnetite [1], Morin transition in hematite [2], low temperature magnetic transition in epsilon ferrite [3], etc. Besides the fundamental interest, potential applications of iron-containing magnetically ordered materials as functional layers for spintronic devices (e.g., for spin filters and spin injectors) are widely discussed [4]. The versatility of iron oxides is enhanced by numerous crystallographic forms in which they can crystallize: there exist at least four known polymorphs of $Fe_2O_3$, namely $\alpha$-$Fe_2O_3$ (hematite), $\beta$-$Fe_2O_3$, $\gamma$-$Fe_2O_3$ (maghemite), and $\varepsilon$-$Fe_2O_3$, as well as $Fe_3O_4$ (magnetite) and $FeO$ (wustite). Of all the iron oxides, the orthorhombic $\varepsilon$-$Fe_2O_3$ phase remains the least studied up to now and does not exist in the bulk form due to its structural metastability. Having attracted a lot of attention during past years, epsilon ferrite in the form of nanofilms [5–7] and nanoparticles [3] was shown to exhibit a complicated magnetic structure with four magnetic sub-lattices, a very high magneto-crystalline anisotropy and room-temperature multiferroic behavior [8] not observed in the other simple metal oxides. The large magnetic coercivity and proven magneto-electric coupling make this material perspective for creation of novel iron-oxide based ferroic-on-semiconductor devices for spintronic applications including low power consumption magnetic media storage devices [9]. Furthermore, epitaxial growth of a room-temperature multiferroic layer with a controllable magnetization/polarization on a semiconductor provides promising opportunities to control optical, electronic, and magnetic properties of such heterostructure by external electric and magnetic fields [10–13]. Remarkably, the majority of the earlier works on metastable $\varepsilon$-$Fe_2O_3$ deal with randomly oriented nanoparticles [9,14,15]. Besides, there are several studies reporting stabilization of $\varepsilon$-$Fe_2O_3$ epitaxial layers on STO, $Al_2O_3$ and YSZ [5,8,16,17]. In the series of recent experiments, the possibility to grow highly ordered layers of $\varepsilon$-$Fe_2O_3$ along with other ($\alpha$-$Fe_2O_3$, $\gamma$-$Fe_2O_3$ and $Fe_3O_4$) iron oxides on top of the GaN (0001) substrate (with and without a MgO buffer layer [18]) by pulsed laser deposition (PLD) has been demonstrated [6]. As it was shown, the polymorphism of iron oxide epitaxial films can be effectively guided by fine tuning technological parameters at deposition stage. The magnetic properties of the $\varepsilon$-$Fe_2O_3$ / GaN films have been described in detail in Refs. [7,19].

The present report adds a detailed in-depth resolution to the structural analysis of the PLD grown epitaxial $\varepsilon$-$Fe_2O_3$ / GaN films. For the first time, high-resolution transmission electron microscopy (HRTEM) has been applied to obtain a detailed depth resolved information on crystal structure, density and chemical composition of epitaxial $\varepsilon$-$Fe_2O_3$ / GaN films. Special attention has been paid to the properties of the interfacial region responsible for nucleation of the metastable epsilon ferrite phase. Understanding crystallographic order and defect structure at the oxide-nitride interface is important since these properties are of critical influence on the performance of potential iron-oxide based ferroic-on-semiconductor devices. The results obtained with electron microscopy study are complemented with high energy electron diffraction (RHEED), polarized neutron reflectometry (PNR) and energy-dispersive X-ray spectroscopy (EDX).

## 2. EXPERIMENTAL

Epitaxial films of $\varepsilon$-Fe$_2$O$_3$ iron oxide were grown on GaN (0001) surface by means of PLD following the routine described in detail in [6,19]. The growth was performed using the PLD system produced by Surface (Germany). Iron oxide was deposited from a stoichiometric Fe$_2$O$_3$ target in the oxygen atmosphere at a pressure of 0.2 mbar onto the GaN substrate heated up to 830 °C. One of the samples fabricated for neutron studies was grown following the same $\varepsilon$-Fe$_2$O$_3$ growth technique but while keeping the substrate at a much lower temperature of 650 °C. Evaporation of the target material was carried out by means of excimer KrF-laser (CompexPro 201 – Coherent, USA, $\lambda$=248 nm). The crystal structure of the films was investigated *in-situ* by reflection high-energy electron diffraction (RHEED) 3D reciprocal space mapping. This method allows obtaining three-dimensional reciprocal space maps from a sequence of diffraction patterns taken during a fine-step azimuthal rotation of the sample [20]. Such 3D-map consists of a dense stack of Ewald spherical sections and can be conveniently analyzed by performing planar cuts perpendicular to the chosen crystallographic axes. The method was proved to be equally effective when applied to in-situ electron diffraction and ex-situ x-ray diffraction studies of nanoscale epitaxial heterostructures [21–23]. Electron microscopy studies were carried out using a FEI Tecnai G2 F20 HRTEM coupled to an EDAX super ultra-thin window (SUTW) X-ray detector. Crystal structure of the film cross-section was examined by HRTEM. Sample cross-sections for HRTEM studies were prepared by standard methods involving: cutting and polishing down to the thickness of 30 microns, ion sputtering (Gatan PIPS) with 4 keV argon ions down to the thickness transparent for TEM, final mild sputtering with 2 keV argon ions at a grazing incidence of 10 degrees. The composition and density depth profiles of $\varepsilon$-Fe$_2$O$_3$ films were studied post-growth by scanning transmission electron microscopy (STEM) operated in high-angle annular dark-field (HAADF) imaging mode. Energy-dispersive X-ray spectroscopy (EDX) measurements were carried out to determine chemical composition of the films, interface sharpness and element mixing. PNR experiments were performed at the SHARAKU time-of-flight instrument (beamline BL17) at J-PARC MLF (Materials and Life Science Experimental Facility) (Japan) [24]. The temperature and magnetic field at the SHARAKU time-of-flight instrument were controlled by a 4 T horizontal field cryomagnet (Cryogenic Ltd, UK). The polarization of the direct beam was measured by the analyzer for each value of the magnetic field used in the experiment and the amplitudes of the PNR curves were corrected for polarization efficiency. For quantitative discussions of the structure and magnetic properties of the studied samples the PNR data was fitted using the Parratt algorithm by the GenX software package [19].

## 3. RESULTS AND DISCUSSION

### 3.1. TEM and EDX-STEM composition profiling

Side-view cross-sections of the epitaxial $\varepsilon$-Fe$_2$O$_3$ / GaN films have been examined by TEM to get a detailed depth-resolved information on the crystal structure. Figure 1 shows a typical medium-resolution TEM image with clearly distinguishable regions belonging to the GaN substrate (bottom) and 50 nm thick $\varepsilon$-Fe$_2$O$_3$ film (top). The estimated peak-to-valley surface roughness of $\varepsilon$-Fe$_2$O$_3$ layer of approximately 10 nm is in good agreement with the previous atomic force microscopy and transmission electron microscopy studies [6]. The roughness is likely caused by the difference in the height of the neighboring columns. Remarkably, as shown in Figure 1, there is a ca. 5 nm transition region at the Fe$_2$O$_3$ / GaN interface with a roughness of less than 1 nm. Figure 2 (b) shows EDX composition profiles of Fe, O, Ga elements measured perpendicularly to the film surface. For a clearer comparison the profiles have been normalized and superimposed onto the STEM image shown in Figure 2 (a). The deficiency of all four elements (Fe, O, Ga, N) in the transition layer is clearly visible from the HAADF STEM signal indicating the overall density decrease compared to the bulk of the film. Moreover, the gallium concentration curve (orange) is seen to penetrate by about 10 nm into the film, thus directly proving the diffusion of gallium from the substrate into the iron oxide layer by the microscopic STEM probe. This diffusion is supposed to have a thermally activated origin since GaN is known to start actively losing gallium at 900 °C [25–27] which is quite close to the growth temperature of 830 °C used in the present work. Noteworthy, the presence of a similar low density transition layer at the $\varepsilon$-Fe$_2$O$_3$ / GaN interface was observed in Ref. [7] and was claimed to be magnetically softer than the main volume of the film. According to the secondary ion mass-spectrometry (SIMS) studies reported in the same paper [7] the transition layer has a GaFeO$_3$-like composition (Ga rich, Fe deficient) and is likely to be formed at the early deposition stage due to thermal migration of Ga atoms from the substrate. Remarkably, the $\varepsilon$-Fe$_2$O$_3$ films discussed in Ref. [7] were grown along the same growth scenario, however the substrate temperature was lower - 700 °C as compared to 830 °C in the present work. Reasonably, at lower growth temperature, the density depression at the interface observed in Ref. [7] by SIMS and PNR was few times less pronounced being narrower and less deep. Interestingly, a similar diffusion of gallium from the substrate was observed earlier by SIMS during epitaxial growth of Y$_3$Fe$_5$O$_{12}$ on Gd$_3$Fe$_5$O$_{12}$ [28] at 700-850 °C. While the EDX data for the sample grown at 650°C is not available, we expect that Ga diffusion is much less pronounced in this sample because of the lower growth temperature.

### 3.2. Depth profiling by PNR

A similar density profile with a dip at the interface was observed in multiple $\varepsilon$-Fe$_2$O$_3$ samples by PNR as shown in Figure 3. Typical reflectivity curves and the corresponding scattering length density

(SLD) depth profiles are shown for a selection of $\varepsilon$-Fe$_2$O$_3$ films. Representative PNR datasets were measured from several $\varepsilon$-Fe$_2$O$_3$/GaN films: two samples (A, B) grown at typical conditions at T = 830 °C and having different thickness of 85 nm and 100 nm, and sample C grown at a much lower temperature of 650°C and with a thickness of 80 nm. The PNR data shown in Figure 3 is multiplied by the forth power of momentum transfer $Q_z^4$ to compensate for the Fresnel decay and to enhance the important high-Q features of the reflectivity curves. Two neutron polarization components of the PNR $R^+(Q_z)$ and $R^-(Q_z)$ parallel and anti-parallel to the saturation magnetic field of 3 T, respectively, were measured at temperatures 300 K and 5 K (shown in Figures 3 (a-c)). The splitting of the $R^+(Q_z)$ and $R^-(Q_z)$ curves is proportional to the net magnetization of the film along the field direction. Because of the relatively small magnetic moment of $\varepsilon$-Fe$_2$O$_3$ the splitting is hardly distinguishable at 300 K but is clearly observed after the sample is cooled down to 5 K in a magnetic field of 3 T.

The PNR data was fitted assuming that the model consists of the GaN substrate, the transition layer (interface) and the main $\varepsilon$-Fe$_2$O$_3$ layer. The resultant models are summarized in the nuclear ($\rho_n$) and magnetic ($\rho_m$) SLD profiles shown in Figure 3 (d). While nuclear SLDs of the 'bulk' layer of samples A and B are identical (Figure 3 (d)), the sample C shows a somewhat elevated density. It may correspond to the additional fraction of $\alpha$-Fe$_2$O$_3$ in the bulk of the film. The minimum model found by the fitting routine contains an additional interface layer with thickness $d_{int}$ = 37 ± 7 Å, roughness $\sigma_{int}$ = 30 ± 10 Å, and nuclear density $\rho_n$ = (3.26 ± 0.49)·10$^{-6}$ Å$^{-2}$ (sample A); $d_{int}$ = 85 ± 15 Å, $\sigma_{int}$ = 27 ± 8 Å, $\rho_n$ = (3.26 ± 0.49)·10$^{-6}$ Å$^{-2}$ (sample B); $d_{int}$ = 35 ± 9 Å, $\sigma_{int}$ = 28 ± 5 Å, $\rho_n$ = (5.64 ± 0.86)·10$^{-6}$ Å$^{-2}$ (sample C) between the iron oxide and the GaN substrate. The thickness and the low nuclear density of the interfacial layer found by modelling the PNR curves is perfectly consistent with the low-density transition layer observed by the real-space TEM shown in the previous section. PNR is a macroscopic probe that averages layer's parameters over the surface area of the sample (15x15 mm$^2$ in the present case). Hence, the relatively large roughness of the interface layer is explained by the slight waviness of the interface also seen in Figure 1, and slightly inhomogeneous distribution of the substrate temperature during the sample growth. Nevertheless, the nuclear SLD contrast between the bulk and the interface layers is clearly pronounced and PNR data cannot be satisfactory fitted without assuming the transition layer between $\varepsilon$-Fe$_2$O$_3$ and GaN. This model is also justified from the real-space TEM data.

Surprisingly, the magnetic counterpart of the SLD $\rho_m$ of the main $\varepsilon$-Fe$_2$O$_3$ layer significantly varies from sample to sample and is minimal for sample C (and other samples grown in similar conditions), which has the *least* pronounced interfacial dip in the nuclear density profile (Figure 3 (d)). Based on this observation it is concluded that the relatively low growth temperature of the sample C improves the homogeneity of the transition layer, as seen as its nuclear SLD, which is close to the bulk layer value. At the same time, the magnetic SLD of this film is greatly decreased compared to samples

A and B (Figure 3(d)), which point to the emergence of inclusions with low magnetization (possibly α-$Fe_2O_3$) in sample C. The low density of the interface layer in samples A and B cannot be explained by a homogenous layer of any combination of elements present in the samples. The only possible reason of this low *average* nuclear density is the presence of macroscopic hollow regions at the interface. This assumption agrees well with the low-density layer seen in the low resolution TEM images. Interestingly, the low density of the transition layer does not negatively affect the magnetic properties of the films, as it is seen from the magnetic SLD profiles of the samples A and B (Figure 3(d)). Moreover, their magnetizations are somewhat increased compared to the previous study [7].

**3.3. FFT analysis of HRTEM images**

The detailed information on the local crystal structure and peculiarities of epitaxial ε-$Fe_2O_3$ / GaN films grown at standard conditions has been obtained using the advanced analysis of HRTEM images. Figure 4 (a) shows a typical HRTEM image of the ε-$Fe_2O_3$ / GaN interface region with clearly identifiable atomic patterns present in the areas corresponding to the substrate, film and the transition layer. In contrast to the substrate which looks mostly homogeneous, the film shows a number of differently looking HRTEM patterns. This is not surprising considering the three-fold epitaxial relations between the orthorhombic ε-$Fe_2O_3$ (001) and hexagonal GaN (0001). As reported in Ref. [6], the crystal axis of the film oriented parallel to the GaN [1–10] in-plane direction must be either ε-$Fe_2O_3$ [100] or ε-$Fe_2O_3$ [−1-10] or ε-$Fe_2O_3$ [−110]. For the convenience of referencing, in what follows these lattice orientations are denoted A, B and C.

To identify the crystallographic phases and lattice orientations the obtained HRTEM images have been analyzed by comparing region-selective fast Fourier transforms (FFT) to the corresponding reciprocal lattices models. Figures 4 (b, c) show large-area FFT patterns obtained from wide image regions corresponding to the film and the substrate with the superimposed reciprocal lattice models. A perfect match is observed between the FFT pattern of the substrate and the modelled reciprocal lattice nodes for GaN (0001) (Figure 4 (c)). A similarly good agreement is observed between the large-area FFT pattern of the film and the reciprocal lattice model calculated for a superposition of A, B and C lattice orientations of ε-$Fe_2O_3$ (Figure 4 (b)). Remarkably, the FFT maps calculated for the film and the substrate are very similar to the corresponding RHEED patterns taken in the same azimuth during sample growth (Figure 4 (d, e)). The high similarity here is due to the fact that RHEED as well integrates over all present lattice orientations.

The known epitaxial relations between ε-$Fe_2O_3$ and GaN must be taken into account to distinguish between individual ε-$Fe_2O_3$ crystallographic domains. In the imaged plane the three possible lattice orientations of ε-$Fe_2O_3$ (A, B, C as denoted above) result in only two distinctly different HRTEM (and the corresponding FFT) patterns. The A-type lattice orientation (with ε-$Fe_2O_3$ [100] axis normal to the drawing plane) results in the A-type pattern and the indistinguishable B- and C-type orientations (with ε-Fe$_2$O$_3$ [110] and [1-10] axes normal to the drawing plane) result in the BC-type pattern. To highlight the HRTEM image regions corresponding to different crystal structures and orientations a color-enhanced Fourier filtering has been applied (Figure 5). Following the general idea of Fourier filtering the HRTEM images analysis was performed according to the following algorithm i) apply FFT to the raw HRTEM image, ii) zero out all Fourier amplitudes outside the chosen reflections and iii) perform the inverse Fourier transform to obtain the final filtered image. The basic high-pass filter (excluding the circular area around the FFT map center) and low-pass filter (excluding the outer borders of the FFT map) have been additionally applied to suppress the slowly changing background and the high-frequency noise. A series of node filters (keeping amplitudes in rectangular areas around the selected reciprocal lattice nodes) have been applied afterwards to highlight regions with the particular direct space periodicities. The filtered image shown in Figure 5 (b) was obtained from the raw HRTEM image shown in Figure 5 (a) by applying the node filters based on the frames sketched in Figure 5 (c). The blue frame was chosen to highlight the ε-Fe$_2$O$_3$ (0 1 3) planes in type A columns, the red one – to highlight the ε-Fe$_2$O$_3$ (-1 1 3) planes in type B columns and the gray one – to highlight the GaN (0 0 2) planes. The ε-Fe$_2$O$_3$ reflections were deliberately chosen unique for a particular type of lattice orientation (type A or B). The reflection used to highlight the transition layer in green was chosen based on the analysis of the region-selective FFT maps discussed in section 3.5 below. The four filtered images (each one with its own node frame) colored gray, green, red and blue have been combined into a single RGB image, to highlight the substrate, the buffer layer and the two differently oriented ε-Fe$_2$O$_3$ columns with individual colors (Figure 5 (b)). The black regions in the final map correspond to the failure of the main periodicities naturally observed at the borders of the columns.

The performed HRTEM image filtering has resulted in high contrast color maps with clearly distinguishable areas of different crystal orientation and structure. In result, the borders between the adjacent ε-Fe$_2$O$_3$ columns in the filtered image shown in Figure 5 (b) appear much sharper compared to the raw image shown in Figure 5 (a). An important advantage of the discussed filtering technique is the opportunity to observe the overlaid regions with different crystal structures that often occur in transmission electron microscopy. For example, an overlay of the red (type A) and blue (type B) columns, that is present in the film appears in violet in Figure 5 (b). The green + blue = cyan and green + red = yellow overlays are also present. Multiple HRTEM images have been filtered by the technique described above to make sure that the columnar structure shown in Figure 5 is typical for the studied ε-Fe$_2$O$_3$ layer. It must be noted that columnar structures similar to the one observed here often occur in heteroepitaxial systems in which the symmetry of the substrate lattice is higher than that of the film [29,30]. In the present study three different orientations of the ε-Fe$_2$O$_3$ lattice with respect to GaN can occur during the nucleation stage. Moreover, there is a great number of equivalent ways of placing the large surface cell of ε-Fe$_2$O$_3$ (001) over the short period surface of GaN. The columnar structure observed in thicker films thus appears upon growth and coalescence of the adjacent nucleation grains that are not in phase with each other.

### 3.4. Direct space modelling of HRTEM images

The direct space structure of the ε-$Fe_2O_3$ / GaN films was further analyzed by superimposing the corresponding lattice models onto the HRTEM image. The model images were calculated by projecting cationic electron densities onto the image plane (Figure 6). The cationic densities were built by placing spherical Gaussians in the cationic positions of the corresponding lattices (Fe in the iron oxide and Ga in the gallium nitride). The positions were calculated using the data available in the form of CIF files for the corresponding crystal lattices. A crystal slab with a thickness of few unit cells was then constructed and projected onto the image plane. In spite of the model simplicity a very good correspondence between the simulation and the HRTEM-images of A-type ε-$Fe_2O_3$ columns (Figure 6 (a)), C-type ε-$Fe_2O_3$ columns (Figure 6 (b)) and GaN (Figure 6 (c)) has been achieved. Noteworthy, the projection model (though not taking into account the phase shift) gives a much better resemblance to the HRTEM images than the ball-and-stick model that shows only the upper atomic layer. The patterns presented in Figures 6 (a) and 6 (b) are in a good agreement with the epsilon ferrite lattice images reported for ε-$Fe_2O_3$ / STO (111) films in Refs. [5] and [31]. In addition to the reasonable fit achieved in the direct space, a good agreement has been observed between the region-selective FFT images and the reciprocal lattice models of ε-$Fe_2O_3$ and GaN (see FFT panels on the right of Figure 6).

Another series of HRTEM images have been obtained with the sample tilted by 30 deg so that the GaN [110] direction gets oriented perpendicular to the image plane to obtain a more detailed information on the crystallographic structure of the ε-$Fe_2O_3$ columns and the interface region. In this configuration the ε-$Fe_2O_3$ columns of types A, B and C are oriented with [±310] or [010] axes perpendicular to the image plane and become indistinguishable - neither in direct nor in the reciprocal space. Remarkably, the borders between the columns are still visible due to the presence of antiphase boundaries. The HRTEM image fragments corresponding to the ε-$Fe_2O_3$ film and to the GaN substrate are shown in Figure 7. A remarkable agreement is achieved between the HRTEM images and the superimposed direct space lattice models as well as between the corresponding FFT maps and the reciprocal space lattice models.

### 3.5. Structural analysis of the interface area

The crystal structure and composition of the transition layer at the ε-$Fe_2O_3$ / GaN interface remains the most unstudied issue so far. It follows from the previous SIMS [7] and current EDX studies that the transition layer is rich in Ga and deficient in Fe. It is reasonable to assume that this layer consists of some form of iron-gallium oxide that would also explain the reduced magnetization detected by PNR. As shown by the HRTEM measurements (see Figure 8 where the large-scale images obtained at two different sample tilts are presented) the transition layer is rich in defects. These defects including stacking faults and dislocations appear at the interface at the initial growth stage and accommodate the strain induced by different crystal structures of the iron oxide layer and the gallium nitride substrate. The selected defect-free fragments of the transition layer area imaged by HRTEM at two different sample tilts are presented in Figure 9. The analysis of the FFT maps shown aside the HRTEM images allows one to make an assumption about the predominant lattice structure of the transition layer. The best fit is achieved for a cubic spinel structure oriented with the [111] axis upwards and with the [1 1 -2] in-plane axis parallel to GaN [1-10]. The average vertical periodicity of 2.40 Å matches well the 2.405 Å distance between the (111) atomic planes in the γ-$Fe_2O_3$ cubic spinel (lattice constant a = 8.33 Å). The equivalent interplane distances in the other iron oxides are worse matched: 2.424 Å and 2.50 Å between the (111) planes in $Fe_3O_4$, and FeO; 2.368 Å between the (001) planes in ε-$Fe_2O_3$ and 2.291 Å between the (0001) planes in α-$Fe_2O_3$ [6].

As shown in Figure 9 a reasonable agreement is achieved between the modelled projections of the cationic electron density corresponding to γ-$Fe_2O_3$ lattice and the experimental HRTEM images measured with two different tilts. The [11-2]-zone HRTEM images of an iron oxide spinel are known to show a pronounced d = 2.4 + 2.4 Å double-layer periodicity [32–34]. This periodicity corresponds to the different structures of the odd and even (111) iron layers in a cubic spinel. The metal atoms in the odd layers are in purely octahedral coordination while the even layers contain both tetrahedrally and octahedrally coordinated metal atoms. The same double-layer periodicity has been also observed by HRTEM in $MgAl_2O_4$ [35] and $CoFe_2O_4$ [34] cubic spinels. In agreement to these observations the direct space patterns also exhibit this double layer periodicity. Remarkably, in the transition layer there are also regions showing a single-layer periodicity and missing reflections in the corresponding Fourier transform. It must be noted that a similar lack of double layer periodicity has been observed earlier in $NiFe_2O_4$ (111) [36] and in $CoFe_2O_4$ (111) [37] spinel films and was explained by the presence of the antiphase boundaries and dislocations similar to those shown in Figure 8. Thus, it can be concluded that the crystal structure of the transition layer highly resembles that of the gamma iron oxide with some sort of cubic/hexagonal polymorphism (that can be caused by diffusion of gallium). The large number of antiphase boundaries makes the stacking sequence switch frequently between ABCABC and ABAB. One has to take into account that unlike in $Fe_3O_4$, in γ-$Fe_2O_3$ there exist iron vacancies in octahedral sites (8/3 out of 24 sites are vacant) [38]. These vacant sites are mainly located in every second row. The ordering of vacancies that has been the subject of investigations for many decades can be partly responsible for the unnaturally looking HRTEM patterns. It is known that those randomly distributed vacancies result in Fd3m space group like in magnetite. The other possible vacancy ordering result in $P4_332$ space group (typical of $LiFe_5O_8$ spinel) [39] as well as $P4_12_12$ space group with c = 3·a [40,41]. However, in the present case the latter two modifications are unlikely to occur as they would result in multiplied periodicities not observed in HRTEM patterns.

Since, the buffer layer contains gallium according to the obtained EDX data, few more assumptions could be made about the composition of the buffer layer. Taking into account that the growth of the iron oxide starts from heating the GaN substrate in the oxygen atmosphere (830 °C) it is reasonable to assume that the transition layer might consist of one of the gallium oxide polymorphs. It must be noted that similar HRTEM FFT patterns have been observed earlier in Ref. [30] in a somewhat similar system: the interface between orthorhombic k-$Ga_2O_3$ (similar in structure to $\varepsilon$-$Fe_2O_3$) and $Al_2O_3$ substrate was claimed to have a cubic spinel crystal structure of $\gamma$-$Ga_2O_3$ (a = 8.23 Å) [42]. The other known gallium oxide modifications $\alpha$-, $\beta$-, $\delta$-, $\varepsilon$-$Ga_2O_3$ do not match FFT maps and the GaO and $Ga_2O$ do not crystallize into a stable form. The less possible candidate is gallium oxynitride that according to Ref. [43] has a hexagonal lattice with parameters close to those of GaN lattice (a = b = 3.19 Å, c = 5.189 Å). To match the HRTEM image, the GaON lattice must be rotated 90° relative to the GaN substrate which seems unreasonable. Still, another candidate is the $\varepsilon$-$Fe_3N$ compound possessing a hexagonal close packed crystal structure with a = 4.77 Å, c = 4.38 Å [44]. All these compounds, though possess a similar lattice, have the interplanar distance quite different from the d = 4.8 Å observed in the present study and must be ruled out. After consideration of various different candidates for the crystal and chemical structure of the transition layer it was concluded that the most reasonable compound in the present case would be a mixed iron-gallium spinel. Such mixed iron-gallium spinel compounds have been observed earlier, e.g., in Ref. [28], where $Ga_{1.8}Fe_{1.2}O_{3.9}$ (with a = 8.3668 Å) obtained by hydrothermal synthesis had the form of spinel as proved by TEM and powder neutron diffraction studies. The lattice constant in a mixed spinel would be somewhere between a = 8.23 Å in pure $\gamma$-$Ga_2O_3$ and a = 8.33 Å in pure $\gamma$-$Fe_2O_3$ in reasonable agreement with the obtained direct space HRTEM images and the corresponding reciprocal space FFT patterns.

## 4. CONCLUSION

In the present work the transverse crystal structure of epitaxial $\varepsilon$-$Fe_2O_3$ films grown on GaN was analyzed for the first time by applying transmission electron microscopy, high energy electron diffraction and polarized neutron reflectometry. The analysis has been carried out simultaneously in the direct and reciprocal spaces by comparing the fast Fourier transform of the HRTEM images to the in situ RHEED patterns as well as by superimposing the modelled crystal lattices (cationic density projections) onto the HRTEM images. According to EDX data, the stoichiometry of the main body of the film above the transition layer is close to $Fe_2O_3$ exhibiting some iron deficiency. This stoichiometry means that all iron is in 3+ state which is consistent with the previous X-ray absorption studies of $\varepsilon$-$Fe_2O_3$ / GaN films [6]. Noteworthy, the iron-rich film with some iron being present in 2+ state would show a cubic (like in FeO and $Fe_3O_4$) rather than orthorhombic crystal structure. It has been confirmed that the film is formed by three types of columns related to three different $\varepsilon$-$Fe_2O_3$ lattice orientations at 120° to each other. A color enhanced FFT node filtering was applied to effectively distinguish the image areas belonging to the different columns and to estimate the size of these columns. The width of the columns was found to be of the order of 10 nm. Unlike the previous studies of the ε-$Fe_2O_3$ / GaN films that were mainly performed by the volume integrating techniques such as x-ray and electron diffraction, the local crystal structure of the epsilon ferrite films was investigated with a high spatial resolution (column-wise) in the present work paying particular attention to the composition of the transition layer at the interface between GaN and ε-$Fe_2O_3$. By obtaining complementary HRTEM images at two different sample tilts it has been found that the crystal structure of the interface layer resembles that of a defective γ-$Fe_2O_3$ spinel. The high defect density presumably appears in order to accommodate the stress arising at the junction of two dissimilar materials such as gallium nitride and iron oxide. The analysis of EDX profiles has shown that gallium diffusion from the substrate into the film occurs during the growth. Thus, it is concluded that the compound at the interface is a mixed iron-gallium spinel. As shown by the neutron reflectometry studies this layer has a lower density and a reduced magnetization. Being not as magnetically hard as the orthorhombic iron-gallium oxide, this layer might account for the soft magnetic component appearing in the magnetization loops of the otherwise magnetically hard epsilon ferrite. The PNR studies have proved that while at a lower growth temperature of 650°C the density depression at the interface is much less pronounced, the overall magnetization of the ε-$Fe_2O_3$ film becomes considerably lower. The high-resolution TEM images of the transition layer shown for the first time in the present work are supposed to be important for the development of ε-$Fe_2O_3$ / GaN heterostructures that can potentially become part of the ferroic-on-semiconductor spintronic devices. As shown in the present paper, the defects related to the nitride-oxide mismatch are mostly accommodated by the transition layer while the bulk of the ε-$Fe_2O_3$ layer grows with the native iron oxide lattice structure. Importantly, the huge magneto-crystalline anisotropy which is the most crucial property of ε-$Fe_2O_3$ is present in the epitaxial ε-$Fe_2O_3$ / GaN films despite the large lattice mismatch observed in this system [19].

**ACKNOWLEDGEMENTS**

The authors wish to acknowledge V. V. Lundin for providing GaN / $Al_2O_3$ wafers. The electron microscopy measurements have been supported through EU's Framework Program Project No. 654360 Nanoscience Foundries and Fine Analysis—Europe (NFFA Europe), NFFA ID 820. Sample preparation for TEM has been performed on the equipment of "Material science and characterization in advanced technology" Federal Joint Research Center supported by the Ministry of Education and Science of the Russian Federation (id RFMEFI62119X0021). The PNR measurements at the Materials and Life Science Experimental Facility of the J-PARC were performed under a user program (Proposal No. A0199). The part of the study related to neutron reflectometry was supported by SNSF Projects No. 200021_188707 and Sinergia CRSII5-171003 Nano Skyrmionics. ICN2 acknowledges financial support from the Spanish Ministry of Economy and Competitiveness, through the "Severo Ochoa"

Programme for Centres of Excellence in R&D (SEV20170706). BB acknowledges funding from Generalitat de Catalunya SGR 327. Sergey Suturin and Polina Dvortsova acknowledge support from the Ministry of Science and Higher Education of the Russian Federation (agreement № 075-15-2021-1349) of the part of the work related to epitaxial growth of ferrimagnetic epsilon ferrite films, their structural characterization by RHEED, and HRTEM image analysis.

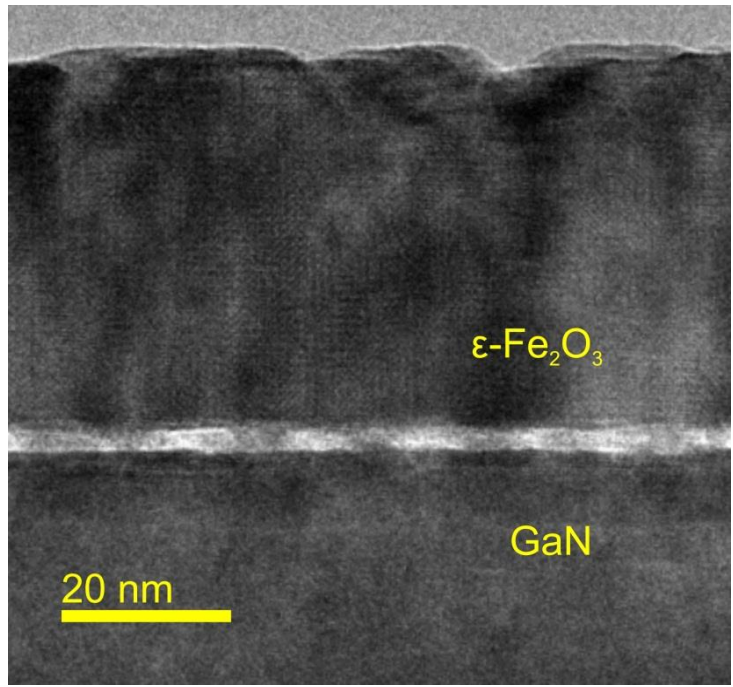

FIG. 1. TEM image of a thin ε-Fe₂O₃ film grown on the GaN (0001) substrate. A few nm thick low-density transition layer can be distinguished at the oxide-nitride interface.

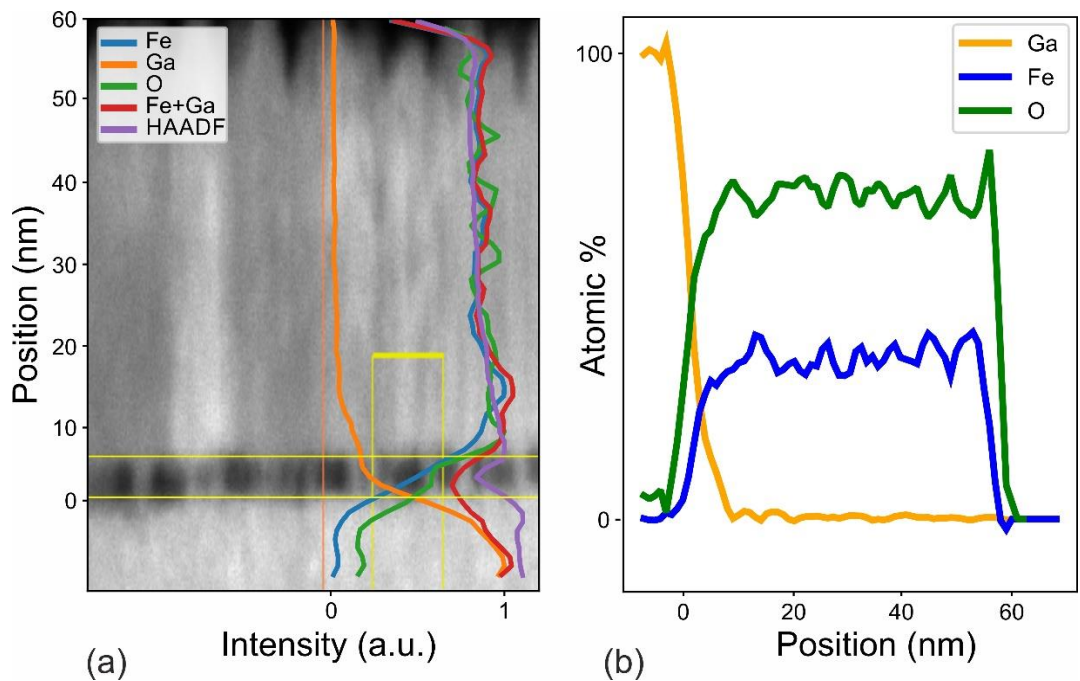

FIG. 2. (A) HAADF STEM image of the ε-Fe₂O₃/GaN interface region with EDX and HAADF profiles superimposed. (B) element selective EDX composition profiles of iron, gallium and oxygen.

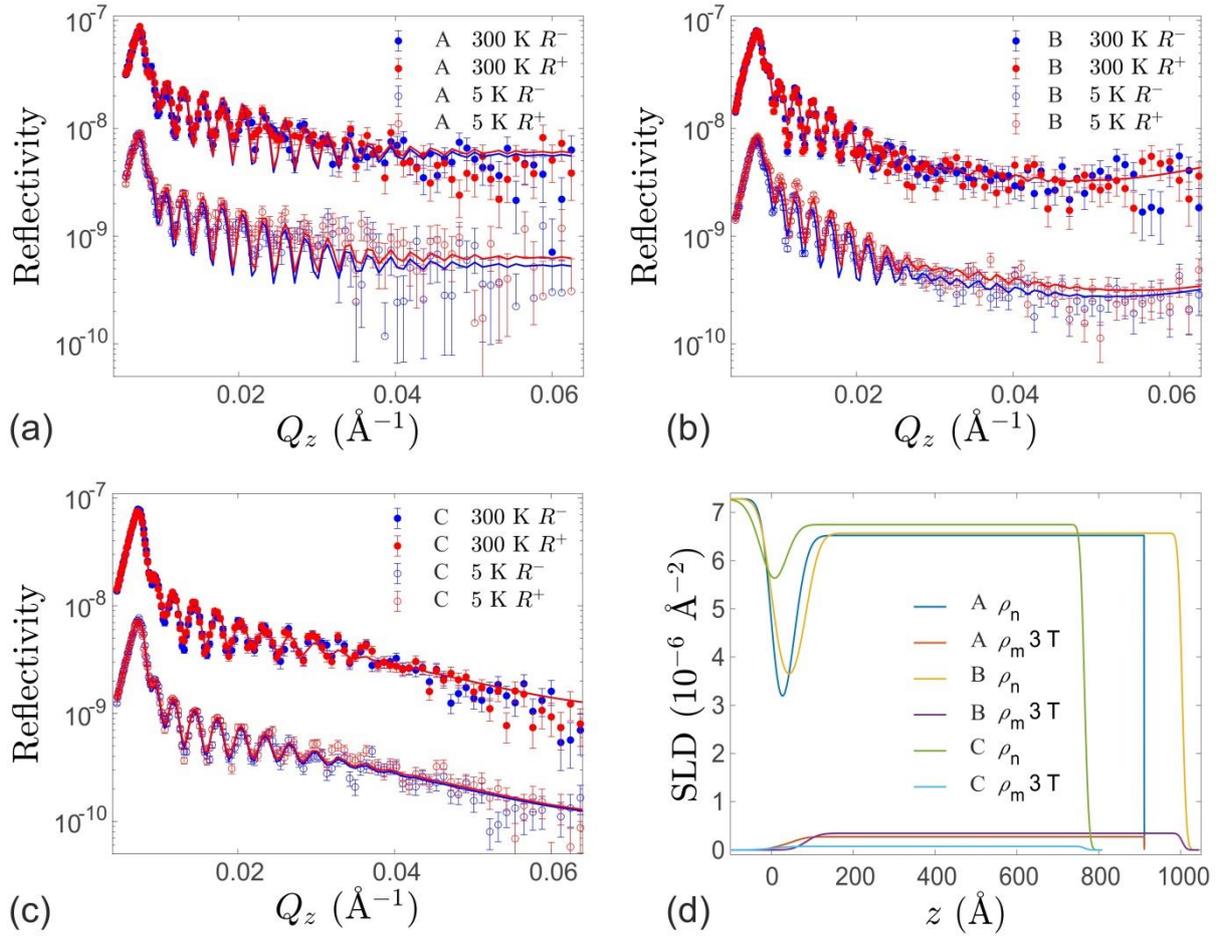

FIG. 3. PNR curves measured in ε-Fe$_2$O$_3$ / GaN samples A (a), B (b) and C (c) in magnetic field of 3 T and at temperatures of 300 K and 5 K. (d) Magnetic and nuclear SLD depth profiles of the films obtained from the fitted model described in the main text. Point z=0 Å corresponds to the ε-Fe$_2$O$_3$ / GaN interface.

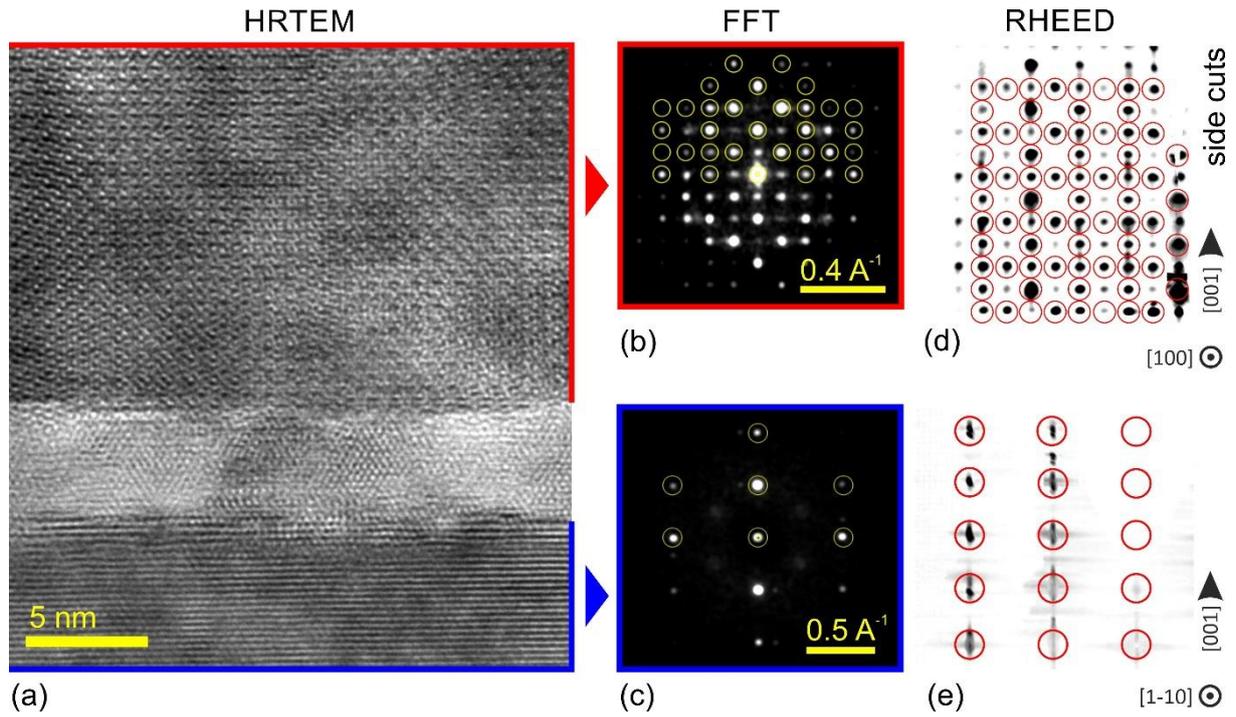

FIG. 4. HRTEM-image of the ε-Fe$_2$O$_3$/GaN interface region taken with the beam parallel to GaN [1-10] axis (a). Large-area FFT maps calculated from the ε-Fe$_2$O$_3$ (b) and GaN (c) image regions. RHEED patterns taken in the same azimuth: during the growth of ε-Fe$_2$O$_3$ film (d) and from the GaN substrate prior to growth (e). The corresponding reciprocal lattices models are superimposed onto the maps shown in panels (b), (c), (d) and (e).

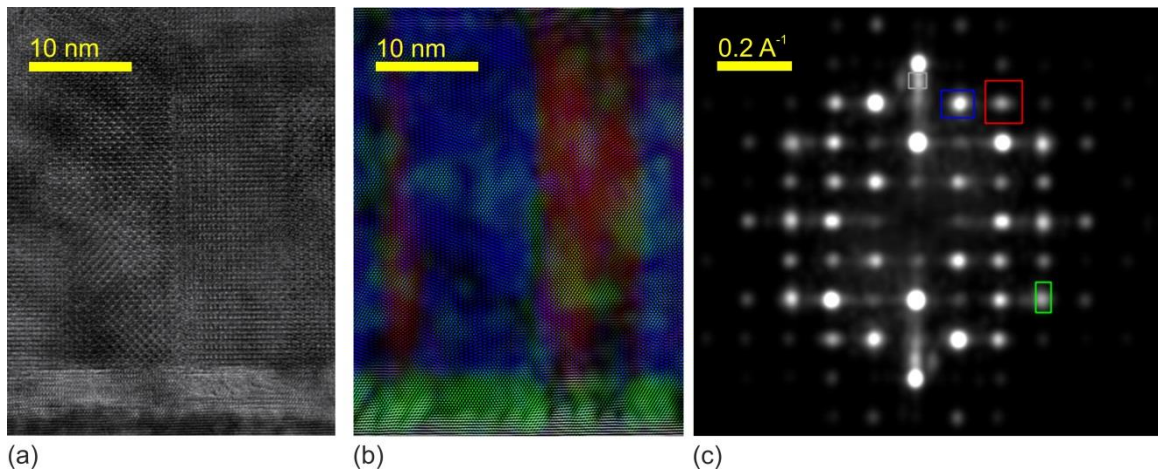

FIG. 5. The raw HRTEM-image of the ε-Fe$_2$O$_3$ / GaN interface region (a). The same image processed by the color-enhanced node filtering (b) with type-A ε-Fe$_2$O$_3$ columns marked red, type-B&C ε-Fe$_2$O$_3$ columns marked blue, GaN substrate marked gray and the transition layer marked green. The nodes selected for node-filtering are shown in the integral FFT map (c) with frames of the same color.

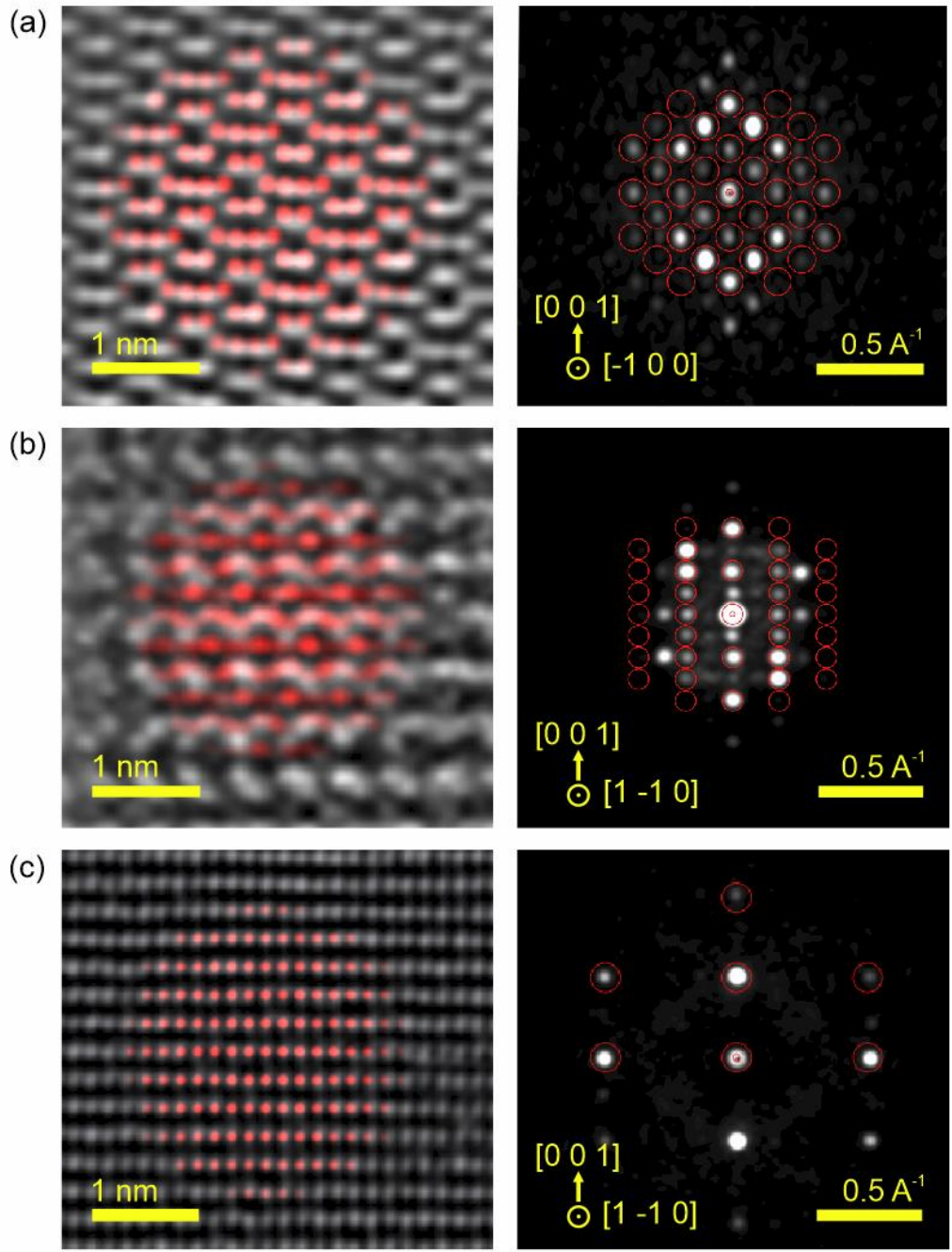

FIG. 6. The HRTEM images with superimposed projection models (left) and the corresponding FFT patterns with overlaid reciprocal lattice models (right) for selected areas in A-type (a) and C-type (b) ε-$Fe_2O_3$ columns and in GaN substrate (c).

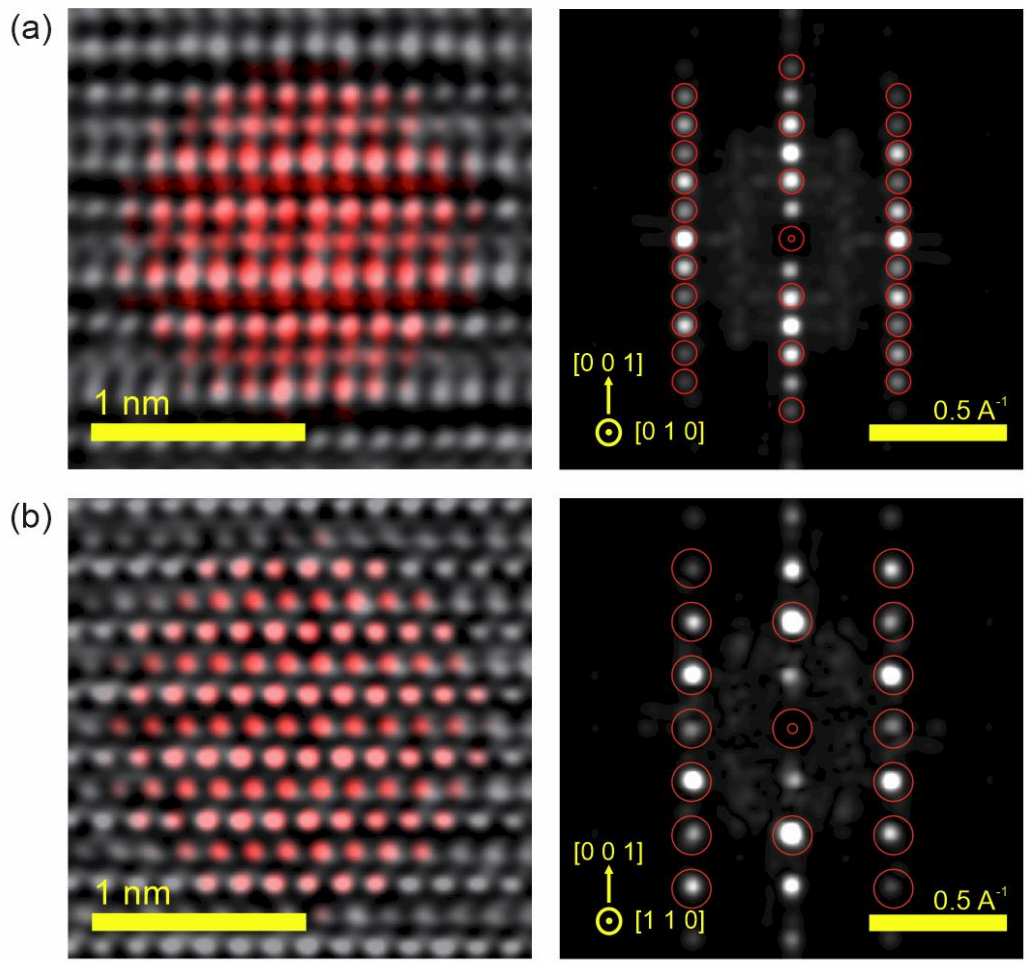

FIG. 7. The HRTEM images with superimposed projection models (left) and the corresponding FFT patterns with overlaid reciprocal lattice models (right) for selected areas in ε-Fe$_2$O$_3$ columns (a) and in GaN substrate (b).

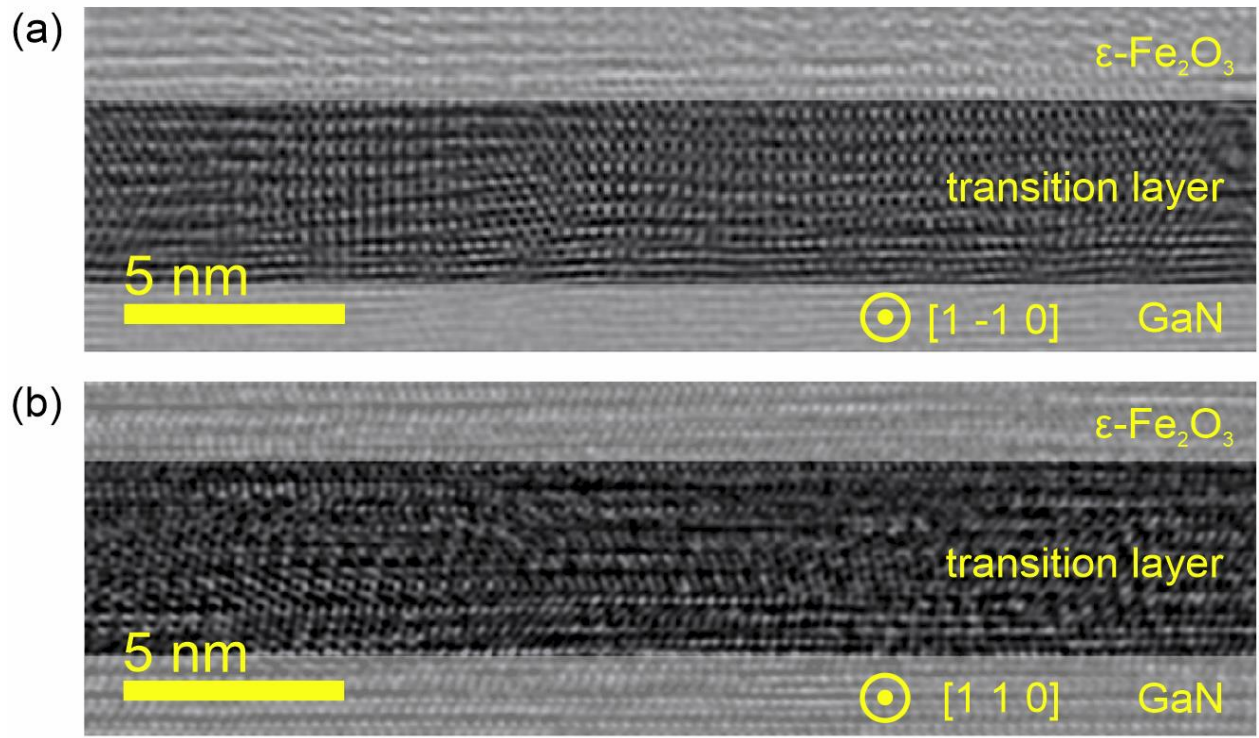

FIG. 8. HRTEM images of the transition layer obtained at two different sample tilts: with GaN [1-10] zone axis (a) and with GaN [110] zone axis (b). The high-frequency noise and low-frequency background have been removed by Fourier filtering.

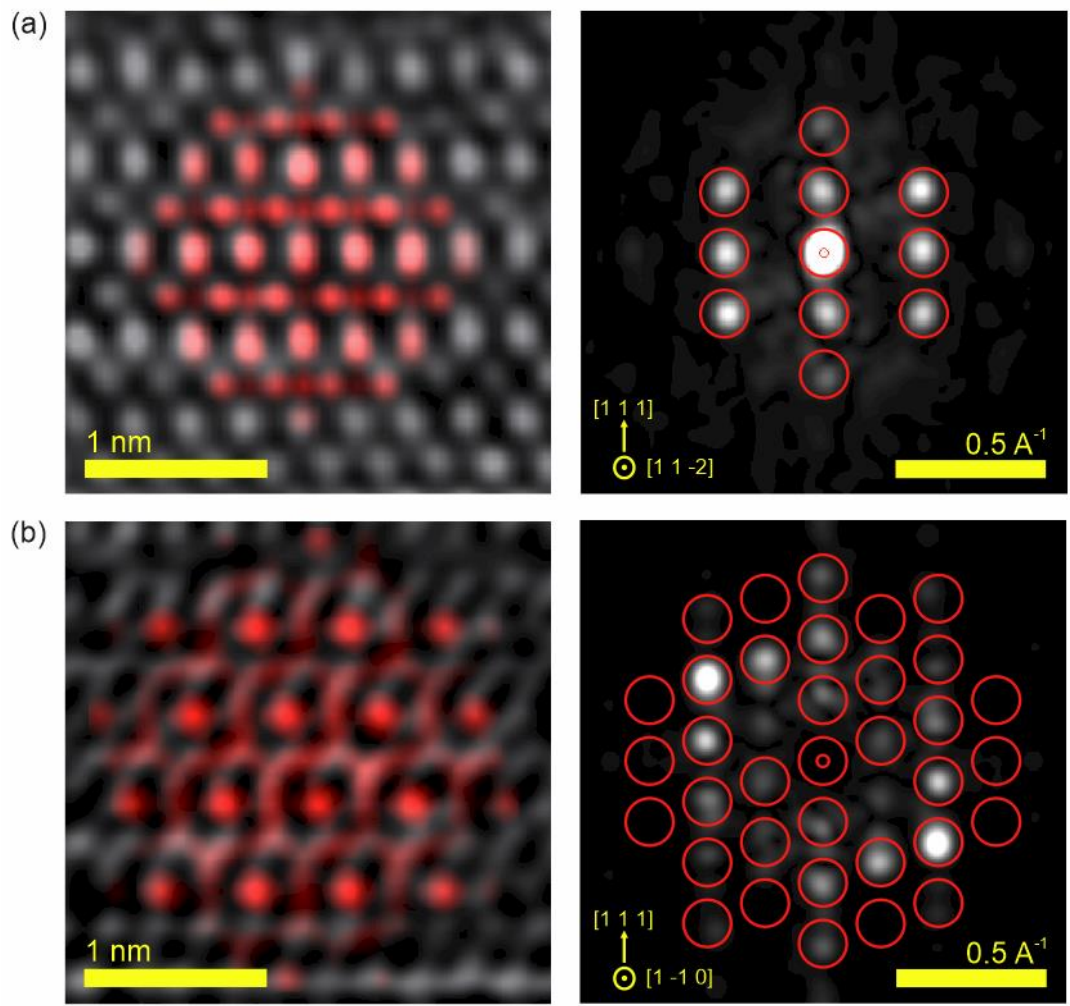

FIG. 9. Defect free fragments of HRTEM images of the transition layer obtained with the two different sample tilts: with GaN [1-10] zone axis (a) and with GaN [110] zone axis (b). The model of the γ-Fe$_2$O$_3$ lattice is superimposed onto the HRTEM images (on the left). The reciprocal lattice nodes corresponding to the γ-Fe$_2$O$_3$ lattice are superimposed on the FFT maps (on the right).